# Visualization of local magnetic moments emerging from impurities in the Hund's metal states of FeSe

Sang Yong Song,[1] J. H. J. Martiny,[2] A. Kreisel,[3] B. M. Andersen,[4] and Jungpil Seo[1,†]

[1]Department of Emerging Materials Science, DGIST, 333 Techno-Jungang-daero, Hyeonpung-Eup, Dalseong-Gun, Daegu 42988, Korea

[2]Center for Nanostructured Graphene (CNG), Department of Physics, Technical University of Denmark, DK-2800 Kongens Lyngby, Denmark

[3]Institut für Theoretische Physik, Universität Leipzig, D-04103 Leipzig, Germany

[4]Niels Bohr Institute, University of Copenhagen, Lyngbyvej 2, DK-2100 Copenhagen, Denmark

**Abstract**

**Understanding the origin of the magnetism of high temperature superconductors is crucial for establishing their unconventional pairing mechanism. Recently, theory predicts that FeSe is close to a magnetic quantum critical point, and thus weak perturbations such as impurities could induce local magnetic moments. To elucidate such quantum instability, we have employed scanning tunneling microscopy and spectroscopy. In particular, we have grown FeSe film on superconducting Pb(111) using molecular beam epitaxy and investigated magnetic excitation caused by impurities in the proximity-induced superconducting gap of FeSe. Our study provides deep insight into the origin of the magnetic ordering of FeSe by showing the way local magnetic moments develop in response to impurities near the magnetic quantum critical point.**

FeSe presents intriguing properties in terms of the interplay among the lattice, charge, and spin degree of freedom. Its nematic phase transition occurs at $T_S$ = 90 K, below which the $C_4$ lattice symmetry is reduced to $C_2$ symmetry [1-3]. Unlike other iron-based superconductors, however, the long-range magnetic ordering is absent in FeSe down to the superconducting transition temperature $T_c$ = 8 K for bulk material, making the degree of freedom that drives the nematic order ambiguous among lattice, charge, and spin [3-10].

Although long-range magnetic ordering is absent in FeSe, there are experiments that suggest the ground state of FeSe is close to the magnetic quantum phase transition point. The hydrostatic pressure of ~1 GPa induces static stripe antiferromagnetic (AFM) orders in FeSe which are typically observed in other iron-based superconductors [11-14]. There is also evidence of local magnetism in FeSe. For example, a muon spin resonance (µSR) study of $FeSe_{0.85}$ measured an exponential decay of the muon polarization, which might hint at the presence of randomly oriented local magnetic moments [15]. The magnetostriction and susceptibility experiment shows strong in-plane anisotropy in FeSe, inferring the coupling of the local magnetic ordering and spin-orbit coupling [16]. Recently, a scanning tunneling microscopy (STM) study observed a signature of local spin fluctuations near the Fe defect in multi-layer FeSe on $SrTiO_3$ substrate [17]. Such magnetic instability suggests the possibility that the magnetism can be triggered by impurities in FeSe [18-20].

Despite intensive efforts in understanding the magnetism in FeSe, the direct observation of local magnetic moments emerging from impurities, which results from the quantum instability, has been challenging mostly owing to the lack of spatial magnetic resolution in the experiments. Here, we report a novel experimental approach to investigate the impurity-induced local magnetic moments in FeSe. Using molecular beam epitaxy (MBE), we grow FeSe film on Pb(111) substrate which is known to be an s-wave superconductor. We have observed a clear signature of the s-wave superconducting gap on the FeSe film, which is proximity-induced from the Pb. When magnetic moments are developed near impurities in FeSe, they respond to the s-wave superconductivity and provide exchange potentials to break Cooper pairs. This leads to strong bound states known as Yu-Shiba-Rusinov (YSR) excitation within superconducting gap [21-25]. Thanks to the extreme sensitivity of superconductivity to magnetism, the energy and spatial resolutions in our study for observing the local magnetic moments are unprecedented. All data are taken at 4.3 K.

In the growth of FeSe, we first grew a single layer (SL) of PbSe on Pb(111). Subsequently, we deposited Fe atoms on the PbSe at 490 K, which resulted in the formation of FeSe islands on Pb(111) (Supplemental Material [26]). Figure 1a shows a typical STM image of the FeSe island grown on Pb(111). The FeSe island is surrounded by the PbSe layer and several bare Pb patches appeared during the growth. The hexagonal-shaped defects in the island and near the island are Ar gas bubbles trapped inside the Pb substrate [30]. The inset shows the atomic structure of FeSe. A rectangular lattice structure is clearly resolved and is distinguished from the crystal structure of Pb(111).

Figure 1b shows the Fourier transform (FT) of the topography of the FeSe. The lattice peaks are present at the position of $(k_x, k_y) = (\pm 1.2, \pm 1.2)$ Å$^{-1}$, whose numbers translate into the lattice constant of 3.7 Å. Figure 1c depicts the atomic model of FeSe forming tri-layer (TL) structure. The blue-filled circles, the red-filled circles, and the blue-open circles represent the Se atoms in the top layer, the Fe atoms in the middle layer and the Se atoms in the bottom layer, respectively. As STM mostly measures the top-most Se atoms [31,32], the obtained lattice structure should conform to the Se lattice in the top layer (sold box in Fig. 1c). For bulk FeSe, the Se lattice constant is known to be ~ 3.75 Å [33-35], which agrees well with the measured value.

The apparent height of FeSe islands with respect to the Pb substrate is found to be ~ 1.7 Å (Fig. 1d). This is smaller than the 1 TL of bulk FeSe (~ 5.33 Å), indicating that most of the FeSe is embedded inside Pb. Similar growth has been reported when FeSe film is grown on soft substrates [36-38]. The lattice modeling of FeSe and Pb estimates the thickness of our FeSe is 3 TL [26]. A Moiré pattern found in the FeSe (Fig. 1a), which is due to the lattice mismatch between FeSe and Pb(111), confirms that the FeSe is in the thin film limit.

To study the electronic property of the FeSe, we performed a differential conductance (dI/dV) spectroscopy using a standard lock-in technique [30]. To maximize the energy resolution in measuring dI/dV spectrum at our experiment temperature, we used a Pb-coated superconducting tip [22,30]. Figure 1e shows the dI/dV spectra measured in the FeSe and Pb. For the FeSe spectrum, there is a characteristic peak near the bias voltage ($V_{bias}$) of -0.3 V, which agrees with the spectra of FeSe in literature [35,39].

When the spectrum is zoomed in around the Fermi energy, a superconducting gap is found (the inset of Fig. 1e). Given by following three facts, we conclude that the superconductivity of our FeSe is proximity-induced from the Pb substrate. First, the superconducting coherence length ($\xi$) of Pb (~830 Å) is much larger than the thickness of the FeSe (~16 Å for 3 TL). Second, the gap is fully developed revealing the s-wave nature in superconductivity. Third, the phonon peaks associated with Pb superconductivity are clearly seen in the spectrum of FeSe (marked by arrows in the inset of Fig. 1e) [40,41]. No hint of unconventional superconductivity is observed. The gap size (4.6 meV) is twice that of the Pb superconducting gap ($2\Delta \approx 2.3$ meV) because the tip is coated with Pb. Under the $T_c$ of Pb superconductivity (~7.2 K), all electrons of FeSe are forced to participate in the proximity-induced s-wave pairing. Any electron pairs which are not in a time reversal symmetry (TRS) relationship will form YSR excitation states within the superconducting gap [21,22,42].

We have investigated the response of FeSe to proximity-induced s-wave superconductivity. Figure 2b shows spectra measured at different FeSe sites marked with A, B, C, and R in Fig. 2a. The spectrum on PbSe (Position R) is first measured as a reference because PbSe is not an intrinsic superconductor and thus its superconductivity is undoubtedly induced by proximity to the Pb. Remarkably, the spectrum measured inside the FeSe island (Position B) exhibits no YSR excitation, indicating that the expected magnetic moment is zero for the ground state. This is microscopic evidence of the absence of static magnetic ordering in FeSe. In contrast, the spectra measured at the boundary of the FeSe island (Position A and C) show strong YSR excitation states, suggesting local magnetic moments are developed along the boundary. The result demonstrates that FeSe is not a conventional non-magnetic material although its ground state preserves TRS [9,12,13,43-45]. Figure 2c displays the dI/dV plot along the dashed line in Fig. 2a. We barely observed a variation in superconductivity inside the FeSe island.

To study the impurity-induced quantum instability in FeSe, we have deposited Ag atoms onto the sample at 20 K (Fig. 3a). The height of Ag atoms is ~ 0.7 Å on the FeSe (Fig. 3b). By careful FT analysis, we determined that the Ag atoms are located on the center of top Se lattice (Fig. 3c and Supplemental Material [26]). Non-magnetic atoms, such as Ag, will not break TRS and thus the s-wave superconductivity should not respond to them. Accordingly, the dI/dV spectrum of the Ag atom on the Pb surface exhibits no YSR excitation (Fig. 3d). We only observed a slight variation in gap size, which might be related to the double Fermi surface of Pb but is not caused by magnetism [46]. By contrast, when we measured the dI/dV spectrum on the Ag atom placed on the FeSe, strong YSR excitation is detected, showing that local magnetic moments are developed. We also observed similar YSR excitation for Au adatoms on the FeSe (Supplemental Material [26]). It is remarkable that such non-magnetic atoms induce local magnetism in FeSe, which supports the assertion that the ground state of FeSe is near a magnetic quantum critical point.

To understand the pattern of local magnetic moments, we measured dI/dV maps at the various energies ($E = eV_{bias}$) of YSR states. Figure 3e shows the topography of a Ag atom and simultaneously obtained dI/dV maps. The most striking feature in the dI/dV maps is the $C_2$ symmetry. The dI/dV patterns at E = -1.85 meV and E = -1.3 meV are split along up-and-down. The dI/dV pattern at E = 1.48 meV is slightly tilted from the up-and-down splitting. These are representative magnetic patterns induced by Ag atoms in FeSe (Supplemental Material [26]). In terms of the reliability of the observed $C_2$ symmetry, the magnetic patterns varied slightly depending on the Ag atom, but the overall $C_2$ symmetry was maintained for the majority of Ag atoms in the repeated experiments (Supplemental Material [26]). We also exclude the Moiré pattern as an origin of the splitting [26].

Several features are present regarding the magnetic patterns. First, the tendency of splitting is the same among the Ag atoms placed on the FeSe in Fig. 3a. No Ag atom showed a splitting along the left-and-right. This symmetry breaking can be attributed to the nematic order in our FeSe. Second, the in-gap states are strongly localized near the Ag atom. We barely observed long-range magnetic ordering near the Ag atom. Third, the magnetic patterns do not follow the full symmetry of the crystal lattice but only satisfy $C_2$ symmetry.

Recent theory predicts that the local magnetic moments in FeSe reflect the momentum structure of the magnetic fluctuations in the bulk [19]. To confirm this, we compared the magnetic patterns with spin models for FeSe. Figure 3f shows the collinear AFM (cAFM) and Néel AFM models. The cAFM model preserves the $C_2$ symmetry of FeSe, which agrees with our magnetic patterns. The Néel AFM model can be ruled out because it does not have $C_2$ symmetry and it has a definite mirror symmetry plane (the dashed line in Fig. 3f) that contradicts the symmetry of the magnetic patterns at E = -1.85 meV and E = -1.3 meV. Therefore, although we have not observed a long-range magnetic ordering near the impurities, the magnetic patterns do reflect the symmetry of the cAFM ordering. Note that the spin angle of 45° in our models is supported by recent µSR measurement [47]. For the spin angle of 0°, however, the magnetic patterns are also consistent with the symmetry of the cAFM ordering (Supplemental Material [26]).

Now we discuss the origin of cAFM order when local magnetic moments are induced by impurities in FeSe. A recent theoretical study, based on a multi-orbital Hubbard model with a band structure relevant for FeSe, mapped out the phase diagram of local impurity-induced magnetism [19]. Importantly, as shown in Ref. [19] the orbital-selectivity characteristic of Hund's metals [33,48] is directly imprinted on the local impurity-induced order, yielding local $(\pi, 0)$ AFM structure versus $(\pi, \pi)$ AFM local order when orbital-selectivity is included or disregarded, respectively. These calculations reveal that strongly anisotropic magnetic fluctuations dictate the detailed structure of induced local magnetic order [19]. A similar transmutation of the structure of the bulk magnetic fluctuations and superconducting pairing, takes place when including orbital-selectivity [29,45,49]. While the results from Ref. [19] focused on impurities centered on Fe sites, we show in Supplemental Material that a Se-centered disorder like Ag also induces local $(\pi, 0)$-structured magnetic order. In addition, we have applied the same theoretical machinery to sample edges, and found that FeSe is very susceptible to induce magnetism strongly localized near the edges (Supplemental Material [26]), in agreement with our STM findings.

We occasionally found a dumbbell-shaped local defect in the FeSe before the deposition of Ag atoms (Fig. 4a). The center of the defect is located at the Fe site as guided by the two dashed lines depicted in Fig. 4a, suggesting it is an Fe vacancy [32,50,51]. When the dI/dV spectrum is measured off the Fe defect, no YSR excitation is observed. However, when it is measured on the defect, strong YSR excitation is observed, indicating that an Fe defect also induces local magnetic moments in FeSe.

Figure 4c shows the topography of the Fe defect and simultaneously measured dI/dV maps. The dashed line depicted represents the mirror symmetry plane imposed by the crystal lattice. The topography is naturally symmetric with respect to this mirror symmetry. Interestingly, the magnetic patterns in the dI/dV maps are not symmetric under the mirror operation [17,48]. Recent theory shows that orbital-selectivity can give rise to chiral patterns in the conductance maps from local magnetic ordering near Fe defects, upon which the mirror symmetry is broken [19]. In fact, the dI/dV map at E = -1.85 meV shows the axis of the magnetic pattern (yellow dotted line) is tilted from the mirror symmetry axis according to the theory. We find that the observed magnetic patterns are again consistent with the symmetry of the cAFM model. Figure 4d shows the cAFM model with the Fe defect. The cAFM ordering directly breaks the mirror symmetry of the crystal lattice. Furthermore, the cAFM ordering breaks the $C_2$ symmetry of FeSe when a defect exists in the Fe site, which is in contrast to the case of the Ag on FeSe. Two green-colored sites in Fig. 4g are then no longer equivalent in terms of symmetry. In the experiment, the magnetic excitation at these sites indeed appears at different energies as shown in Fig. 4c (E = -1.4 meV and E = -0.4 meV).

Our STM experiment provides a novel method to study local magnetism in correlated superconductors, here exemplified through FeSe. The results lead to several important remarks. First, the magnetic quantum phase transition by non-magnetic impurities is microscopically observed in FeSe. Second, we show that the magnetic patterns of the local magnetic moments are consistent with the ($\pi$, 0) AFM phase, implying that the orbital-selectivity is at play. Third, our experiment reveals the magnetic characteristics of impurities in FeSe. The sign-changing superconductivity responds to both magnetic and non-magnetic impurities, whereas the $s_{++}$ superconductivity only responds to magnetic impurities [21-25,52-56]. It is therefore important to characterize the magnetic property of impurities before they are used to probe the symmetry of superconductivity. In our experiment we have unambiguously observed signatures of magnetism on crystalline defects like the crystal boundary and Fe vacancy in FeSe, a property which has not been considered earlier [57,58]. Furthermore, it should be noted that the local magnetism can also be induced by non-magnetic impurities in FeSe.

The data analysis described here is based on a simple and powerful symmetry argument. A theoretical work deserving further investigation is identifying the Fe orbitals responsible for magnetic patterns observed in the experiment. This will reveal the origin of the local magnetism in FeSe in conjunction with the orbital-selectivity. In future STM works, it will be interesting to study how the local magnetism develops into the bulk magnetism when the Ag impurities form networks, which can be accomplished by the STM atom manipulation. Nearby, it might be possible to detect strong orbital-selective spin fluctuations through inelastic tunneling spectroscopy, which could be in turn related to the anisotropic Cooper pairing in FeSe [59].

Acknowledgments

This work has been supported by Samsung Science & Technology Foundation under Project Number SSTF-BA1502-04. The Center for Nanostructured Graphene (CNG) is supported by the Danish National Research Foundation, Project DNRF103. B. M. A. acknowledges support from the Independent Research Fund Denmark, grant number DFF-8021-00047B.

†jseo@dgist.ac.kr

References

[1]   A. I. Coldea and M. D. Watson, Annual Review of Condensed Matter Physics **9**, 125 (2018).

[2]   A. E. Böhmer and A. Kreisel, Journal of Physics: Condensed Matter **30**, 023001 (2018).

[3]   T. M. McQueen, A. J. Williams, P. W. Stephens, J. Tao, Y. Zhu, V. Ksenofontov, F. Casper, C. Felser, and R. J. Cava, Physical Review Letters **103**, 057002 (2009).

[4]   M. Bendele *et al.*, Physical Review Letters **104**, 087003 (2010).

[5]   T. Imai, K. Ahilan, F. L. Ning, T. M. McQueen, and R. J. Cava, Physical Review Letters **102**, 177005 (2009).

[6]   J.-H. She, M. J. Lawler, and E.-A. Kim, Physical Review Letters **121**, 237002 (2018).

[7]   F. Wang, S. A. Kivelson, and D.-H. Lee, Nature Physics **11**, 959 (2015).

[8]   J. K. Glasbrenner, I. I. Mazin, H. O. Jeschke, P. J. Hirschfeld, R. M. Fernandes, and R. Valentí, Nature Physics **11**, 953 (2015).

[9]   R. M. Fernandes, A. V. Chubukov, and J. Schmalian, Nature Physics **10**, 97 (2014).

[10]  S. H. Baek, D. V. Efremov, J. M. Ok, J. S. Kim, J. van den Brink, and B. Büchner, Nature Materials **14**, 210 (2014).

[11]  M. Bendele *et al.*, Physical Review B **85**, 064517 (2012).

[12]  J. P. Sun *et al.*, Nature Communications **7**, 12146 (2016).

[13]  K. Kothapalli *et al.*, Nature Communications **7**, 12728 (2016).

[14] T. Terashima *et al.*, J Phys Soc Jpn **84**, 063701 (2015).

[15] R. Khasanov *et al.*, Physical Review B **78**, 220510(R) (2008).

[16] M. He, L. Wang, F. Hardy, L. Xu, T. Wolf, P. Adelmann, and C. Meingast, Physical Review B **97**, 104107 (2018).

[17] W. Li *et al.*, Nature Physics **13**, 957 (2017).

[18] M. N. Gastiasoro and B. M. Andersen, Journal of Superconductivity and Novel Magnetism **28**, 1321 (2014).

[19] J. H. J. Martiny, A. Kreisel, and B. M. Andersen, Physical Review B **99**, 014509 (2019).

[20] M. N. Gastiasoro, P. J. Hirschfeld, and B. M. Andersen, Physical Review B **88**, 220509(R) (2013).

[21] A. Yazdani, B. A. Jones, C. P. Lutz, M. F. Crommie, and E. D. M., Science **275**, 1767 (1997).

[22] K. J. Franke, G. Schulze, and J. I. Pascual, Science **332**, 940 (2011).

[23] H. Shiba, Progress of Theoretical Physics **40**, 435 (1968).

[24] M. I. Salkola, A. V. Balatsky, and J. R. Schrieffer, Physical Review B **55**, 12648 (1997).

[25] A. V. Balatsky, I. Vekhter, and J.-X. Zhu, Reviews of Modern Physics **78**, 373 (2006).

[26] See Supplemental Material [url] for details, which includes Refs. [19, 27-29].

[27] A. Weiße, G. Wellein, A. Alvermann, and H. Fehske, Reviews of Modern Physics **78**, 275 (2006).

[28] L. Covaci, F. M. Peeters, and M. Berciu, Physical Review Letters **105**, 167006 (2010).

[29] A. Kreisel, B. M. Andersen, P. O. Sprau, A. Kostin, J. C. S. Davis, and P. J. Hirschfeld, Physical Review B **95**, 174504 (2017).

[30] S. Y. Song and J. Seo, Scientific Reports **7**, 12177 (2017).

[31] C. L. Song *et al.*, Science **332**, 1410 (2011).

[32] P. Choubey, T. Berlijn, A. Kreisel, C. Cao, and P. J. Hirschfeld, Physical Review B **90**, 134520 (2014).

[33] A. Kostin, P. O. Sprau, A. Kreisel, Y. X. Chong, A. E. Böhmer, P. C. Canfield, P. J. Hirschfeld, B. M. Andersen, and J. C. S. Davis, Nature Materials **17**, 869 (2018).

[34] F. S. Li *et al.*, 2d Mater **3**, 024002 (2016).

[35] X. Liu *et al.*, Journal of Physics: Condensed Matter **27**, 183201 (2015).

[36] A. Cavallin *et al.*, Surface Science **646**, 72 (2016).

[37] A. Eich *et al.*, Physical Review B **94**, 125437 (2016).

[38] U. R. Singh, J. Warmuth, V. Markmann, J. Wiebe, and R. Wiesendanger, Journal of Physics: Condensed Matter **29**, 025004 (2017).

[39] Y. H. Yuan *et al.*, Nano Letters **18**, 7176 (2018).

[40] M. Schackert, T. Märkl, J. Jandke, M. Hölzer, S. Ostanin, E. K. U. Gross, A. Ernst, and W. Wulfhekel, Physical Review Letters **114**, 047002 (2015).

[41] W. L. McMillan and J. M. Rowell, Physical Review Letters **14**, 108 (1965).

[42] D. J. Choi, C. Rubio-Verdu, J. de Bruijckere, M. M. Ugeda, N. Lorente, and J. I. Pascual, Nature Communications **8**, 15175 (2017).

[43] Q. Wang *et al.*, Nature Communications **7**, 12182 (2016).

[44] Q. Wang *et al.*, Nature Materials **15**, 159 (2015).

[45] T. Chen *et al.*, Nature Materials **18**, 709 (2019).

[46] M. Ruby, B. W. Heinrich, J. I. Pascual, and K. J. Franke, Physical Review Letters **114**, 157001 (2015).

[47] R. Khasanov, Z. Guguchia, A. Amato, E. Morenzoni, X. Dong, F. Zhou, and Z. Zhao, Physical Review B **95**, 180504(R) (2017).

[48] P. O. Sprau *et al.*, Science **357**, 75 (2017).

[49] A. Kreisel, B. M. Andersen, and P. J. Hirschfeld, Physical Review B **98**, 214518 (2018).

[50] S. Chi *et al.*, Physical Review B **94**, 134515 (2016).

[51] D. Huang, T. A. Webb, C.-L. Song, C.-Z. Chang, J. S. Moodera, E. Kaxiras, and J. E. Hoffman, Nano Letters **16**, 4224 (2016).

[52] S. H. Pan, E. W. Hudson, K. M. Lang, H. Eisaki, S. Uchida, and J. C. Davis, Nature **403**, 746 (2000).

[53] H. Yang, Z. Wang, D. Fang, Q. Deng, Q.-H. Wang, Y.-Y. Xiang, Y. Yang, and H.-H. Wen, Nature Communications **4**, 2749 (2013).

[54] Y. J. Yan *et al.*, Physical Review B **94**, 134502 (2016).

[55] Z. Du *et al.*, Nature Physics **14**, 134 (2017).

[56] C. Liu, Z. Wang, Y. Gao, X. Liu, Y. Liu, Q.-H. Wang, and J. Wang, Physical Review Letters **123**, 036801 (2019).

[57] L. Jiao, S. Rößler, C. Koz, U. Schwarz, D. Kasinathan, U. K. Rößler, and S. Wirth, Physical Review B **96**, 094504 (2017).

[58] Z. Ge, C. Yan, H. Zhang, D. Agterberg, M. Weinert, and L. Li, Nano Letters **19**, 2497 (2019).

[59] S. Chi *et al.*, Nature Communications **8**, 15996 (2017).

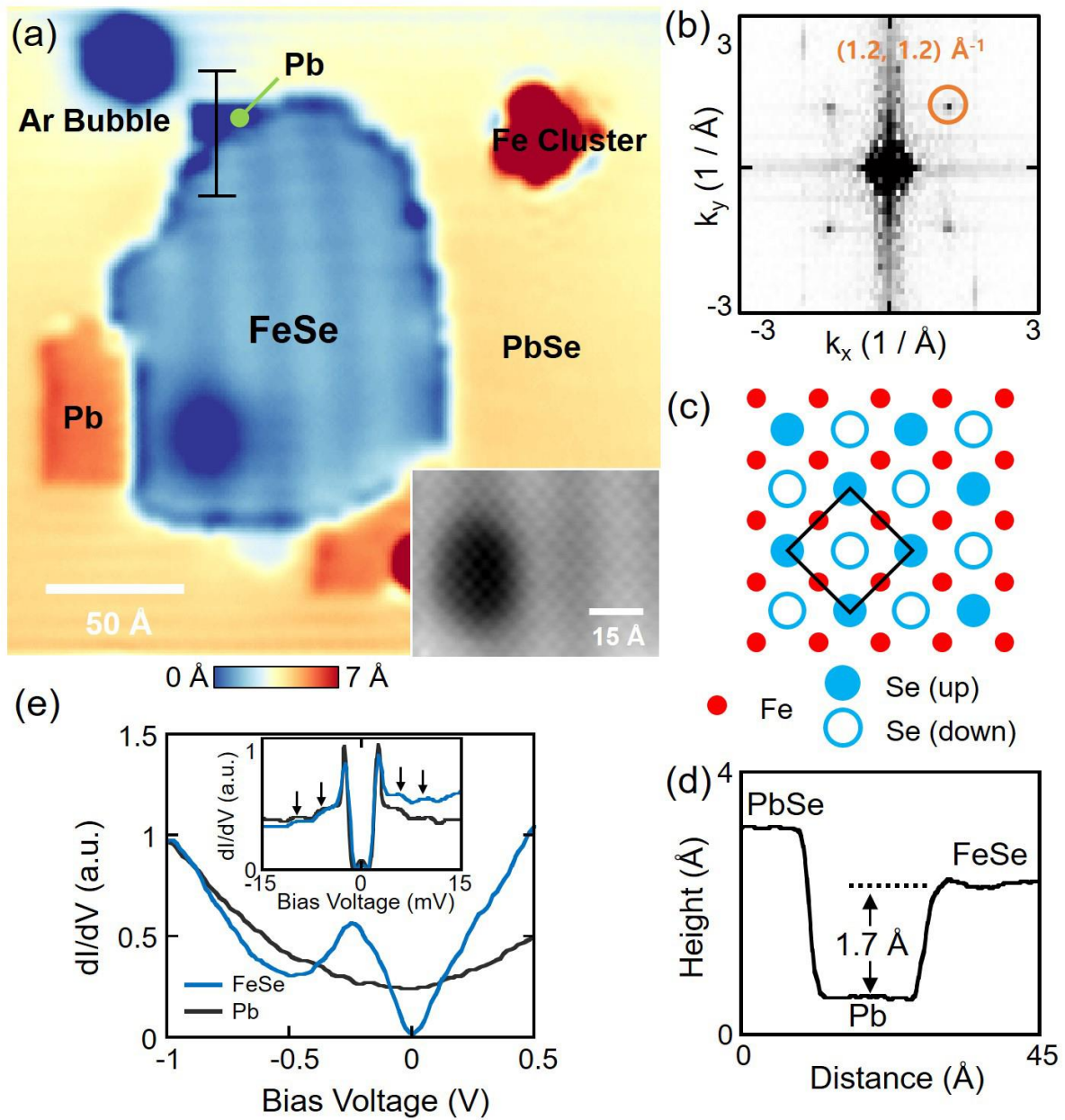

**Figure 1.** (a) The topography of FeSe island grown on Pb(111). $V_{bias}$ = -0.1 V and I = 50 pA. The inset shows the atomically resolved image. (b) Fourier transform of the FeSe image. (c) Atomic model of the FeSe. The solid box represents the Se lattice in the top layer. (d) The height profile along the vertical line in (a). (e) The dI/dV spectra measured in the FeSe island and bare Pb. $V_{bias}$ = -1 V and I = 50 pA. Lock-in: frequency f = 463.0 Hz and root-mean-square (rms) amplitude $V_{rms}$ = 10 mV. The inset shows the spectra around the Fermi energy. $V_{bias}$ = -15 mV and I = 50 pA. Lock-in modulation: $V_{rms}$ = 0.3 mV. The arrows indicate the phonon peaks derived from the Pb superconductivity.

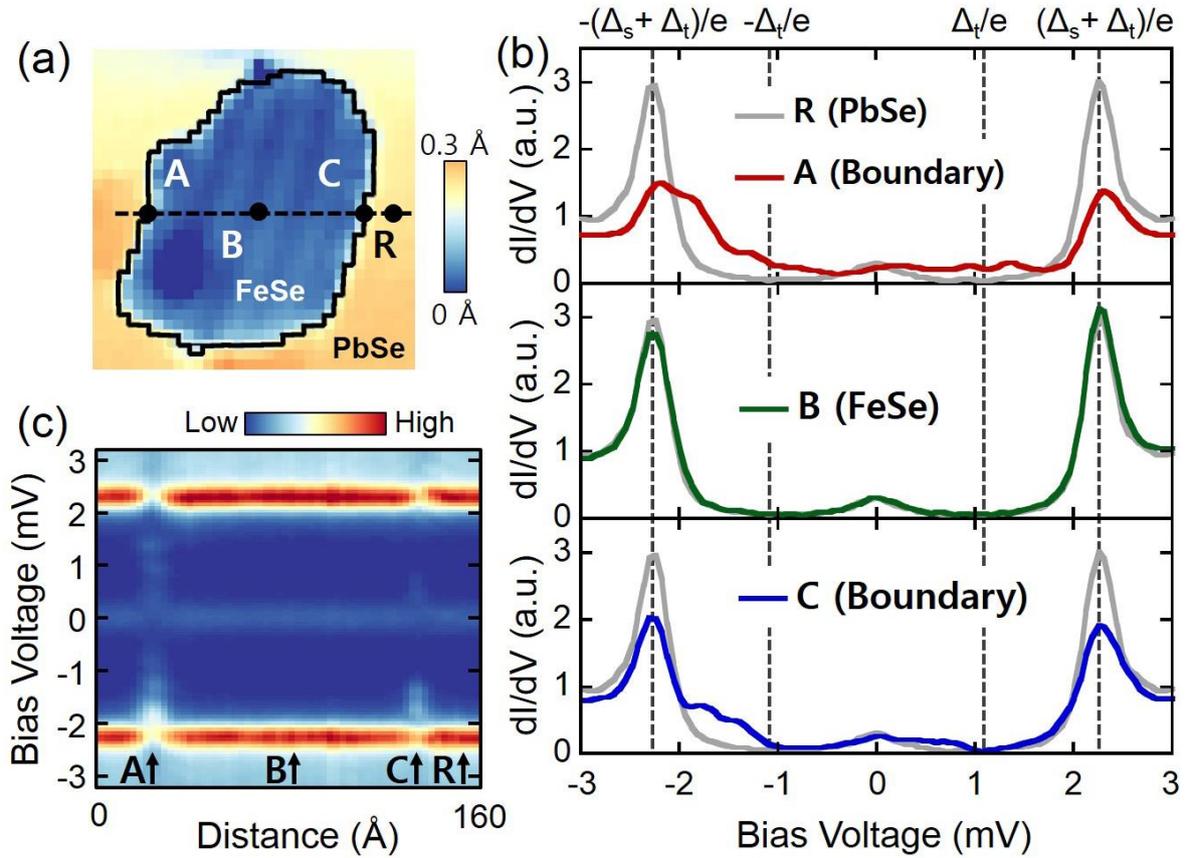

**Figure 2.** (a) The FeSe grown on Pb(111). $V_{bias}$ = -0.1 V and I = 50 pA. (b) The dI/dV spectra measured at Point A, B, C and R marked in (a). $V_{bias}$ = -3.0 mV and I = 50 pA. Lock-in: $V_{rms}$ = 60 μV. Because the Pb-coated superconducting tip is used, the coherent peaks are located at $E = \pm(\Delta_s+\Delta_t)$. $\Delta_s$ and $\Delta_t$ are the superconducting gap of the sample and the tip, respectively. The broad peak around $V_{bias}$ = 0 mV is due to the thermal effect at 4.3 K. (c) Line dI/dV spectroscopy measured along the dashed line in (a).

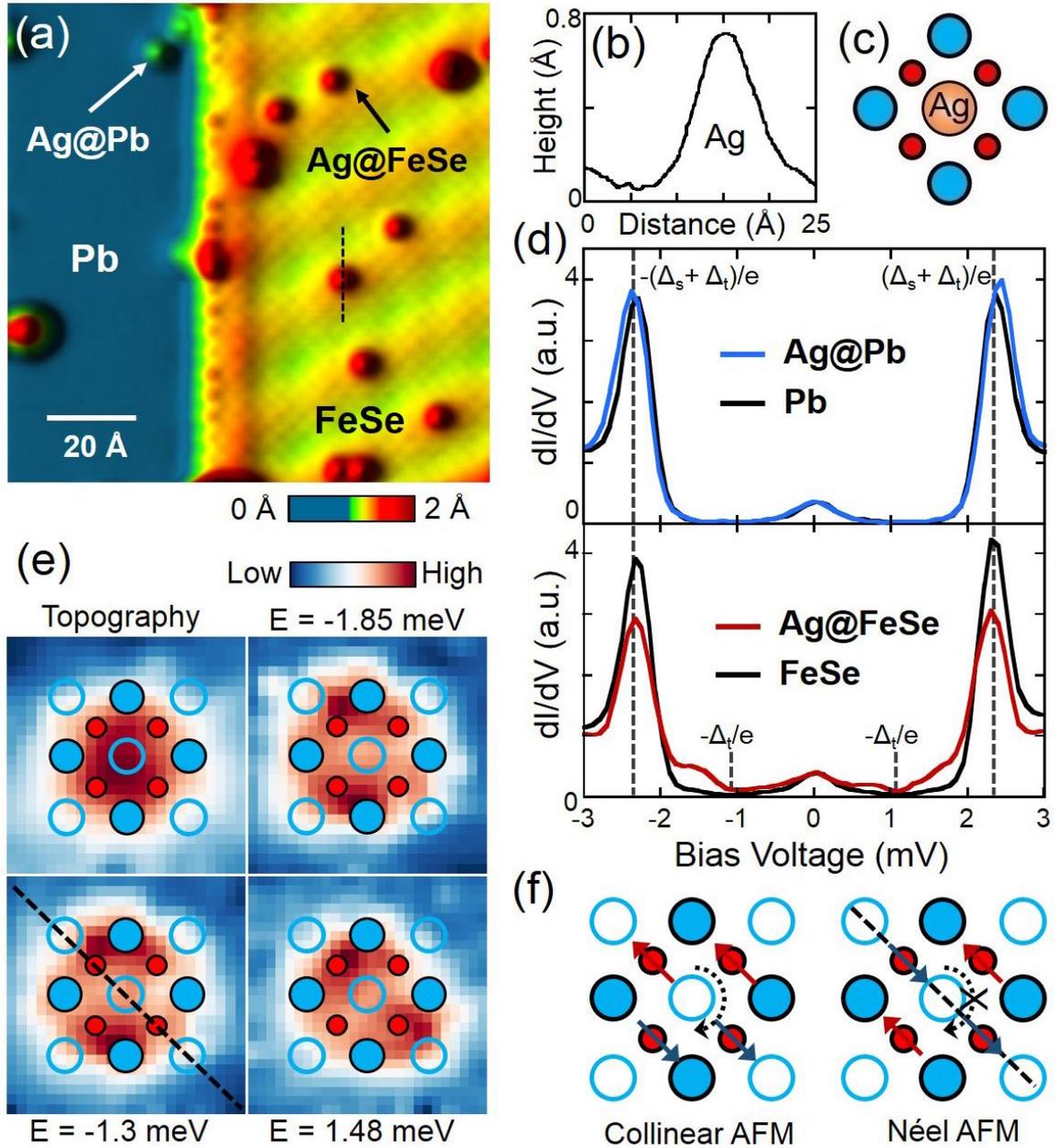

**Figure 3.** (a) Topography of Ag atoms on the FeSe and Pb surface. The Moiré pattern is seen in the FeSe along diagonal direction. $V_{bias}$ = -0.1 V and I = 50 pA. (b) The height profile along the dashed line in (a). (c) The Ag atom (orange ball) is placed on the center of top Se lattice. (d) The Ag atom on Pb do not show in-gap states excitation. By contrast, the Ag atoms on FeSe show strong in-gap states excitation. $V_{bias}$ = -3.0 mV and I = 50 pA. Lock-in: $V_{rms}$ = 60 µV. (e) Topography of the Ag atom and dI/dV maps (size: 9.5 Å x 9.5 Å). (f) Two spin models; collinear AFM model preserves the $C_2$ symmetry imposed by the lattice, as indicated by the dotted arrow. The Néel AFM model breaks the $C_2$ symmetry while it maintains mirror symmetry marked with the dashed line.

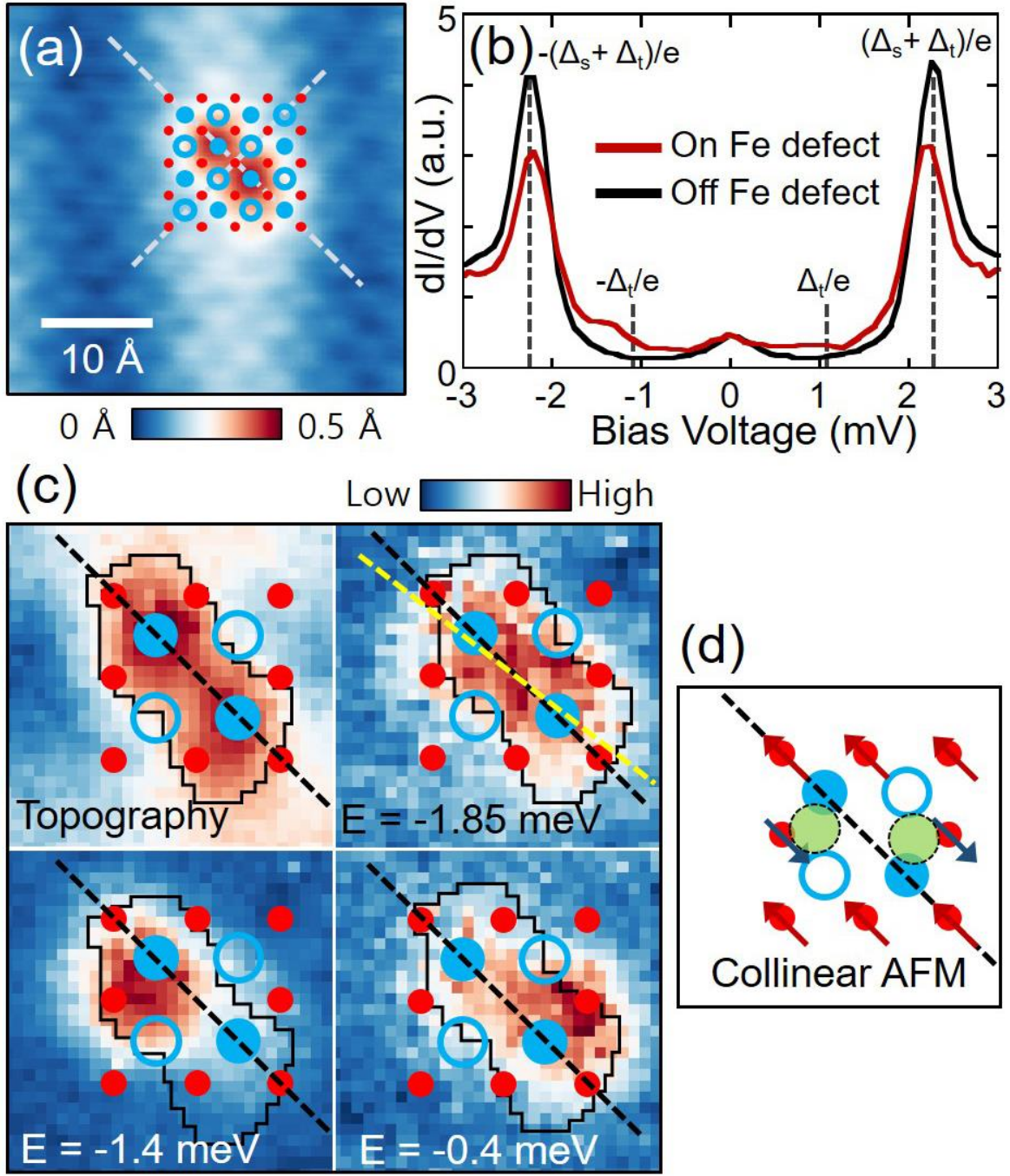

**Figure 4.** (a) Topography of an Fe defect in the FeSe. $V_{bias}$ = -30 mV and I = 50 pA. (b) On the defect site, strong in-gap states excitation is observed. No in-gap states are present off the defect site. $V_{bias}$ = -3.0 mV and I = 50 pA. Lock-in: $V_{rms}$ = 60 μV. (c) Topography of the Fe defect and simultaneously obtained dI/dV maps. The dashed line represents the mirror symmetry plane in topography. The yellow dashed line denotes the axis of the magnetic pattern at E = -1.85 meV. (d) The collinear AFM model breaks the $C_2$ symmetry around the defect as well as the mirror symmetry.

# Supplemental Material

# Visualization of local magnetic moments emerging from impurities in the Hund's metal states of FeSe

Sang Yong Song,[1] J. H. J. Martiny,[2] A. Kreisel,[3] B. M. Andersen,[4] and Jungpil Seo[1,†]

[1]Department of Emerging Materials Science, DGIST, 333 Techno-Jungang-daero, Hyeonpung-Eup, Dalseong-Gun, Daegu 42988, Korea

[2]Center for Nanostructured Graphene (CNG), Department of Physics, Technical University of Denmark, DK-2800 Kongens Lyngby, Denmark

[3]Institut für Theoretische Physik, Universität Leipzig, D-04103 Leipzig, Germany

[4]Niels Bohr Institute, University of Copenhagen, Lyngbyvej 2, DK-2100 Copenhagen, Denmark

†jseo@dgist.ac.kr

Contents

**Note S1. Growth of FeSe on Pb(111) substrate.**

We grew the FeSe film on Pb(111) substrate in a molecular beam epitaxy (MBE) chamber under ultra-high vacuum (UHV) condition. First, the substrate was cleaned by repeated cycles of 2 kV Ar$^+$ sputtering for 10 minutes in Ar pressure of 4.5 x 10$^{-5}$ torr and annealing at 500 K for 12 minutes (min). To form a single layer of PbSe on Pb(111), we evaporated Se atoms for 200 seconds (s) with the speed of ~1 Å/min onto the substrate at 490 K. The sample was then cooled down to 300 K. Subsequently, we heated the sample to 490 K again and Fe atoms were deposited onto the sample for 100 s with the speed of ~1 Å/min. After the Fe deposition, the sample was cooled down to 300 K. The examples of grown FeSe film are provided in Fig. S1.

During the growth of FeSe, Se atoms are supplied from the PbSe layer, which induces the exposure of Pb surface. We observed such Pb surface next to the grown FeSe film and near the step edges of PbSe layer.

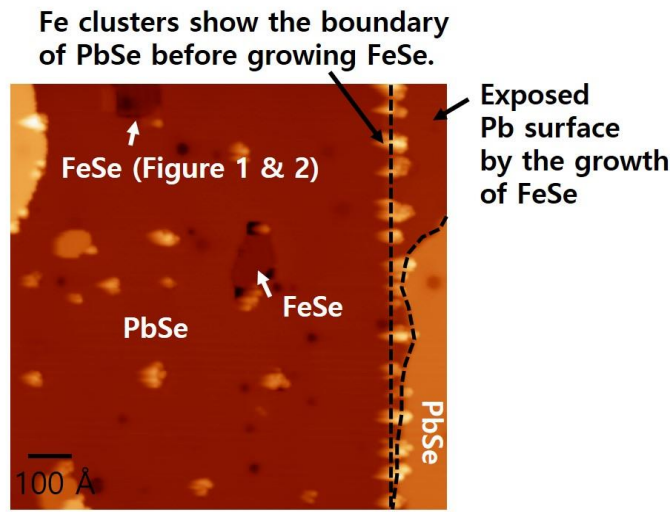

Figure Note S1. The exposure of Pb surface by the growth of FeSe. The growth of FeSe leads to exposure of Pb surface from PbSe because Se atoms needed for the growth are supplied from the PbSe.

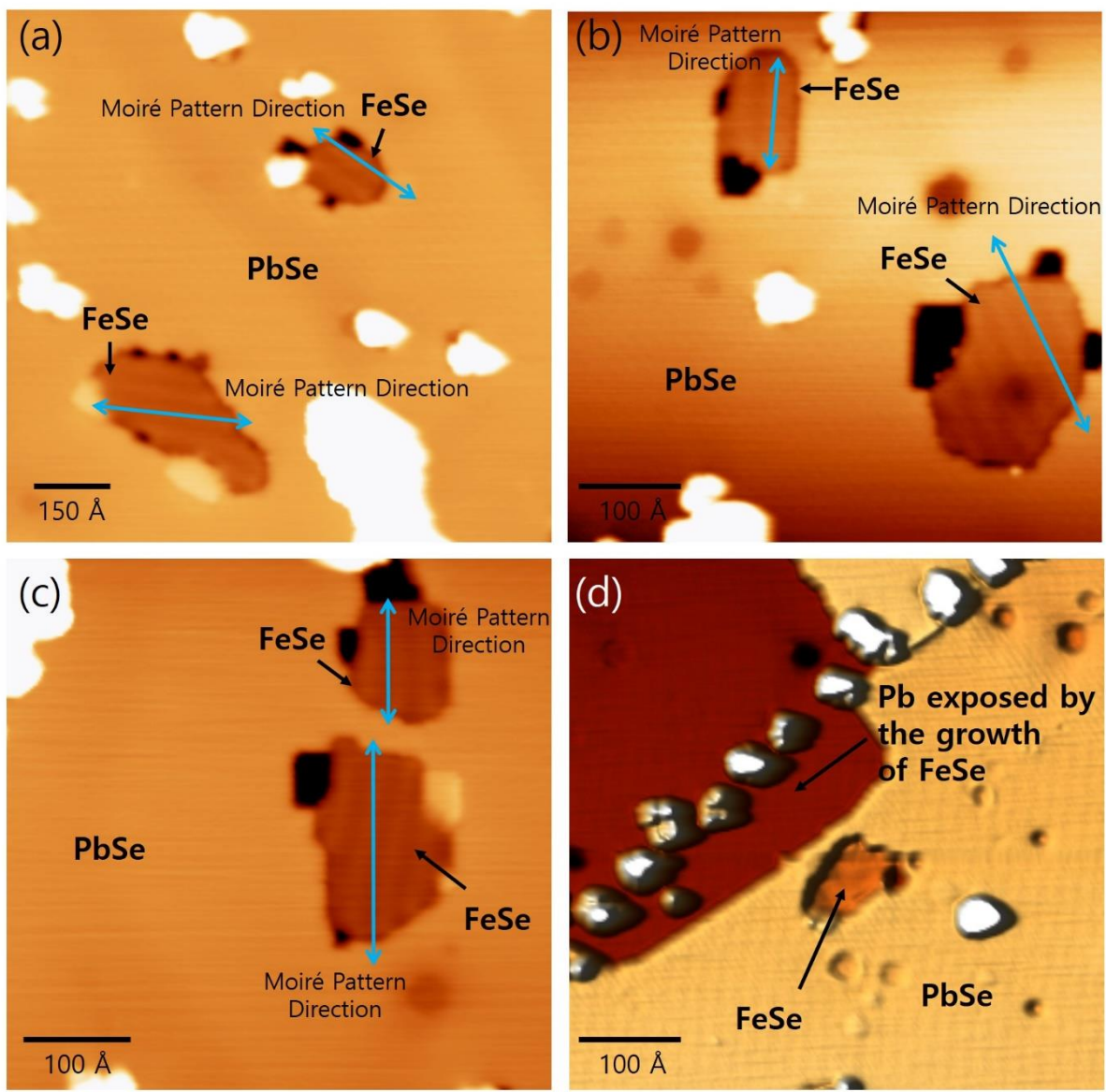

**Figure S1. Examples of FeSe islands grown on Pb(111) substrate.** The size of the grown FeSe islands is typically less than 200 Å x 200 Å. The islands larger than this size was rarely found in our growth condition. (a-c) The Moiré pattern depends on the relative crystal angle between the FeSe layer and Pb substrate. (d) A 3-dimensional rendered image of FeSe/PbSe/Pb(111). The growth of FeSe induced the exposure of Pb surface near the step edge of PbSe.

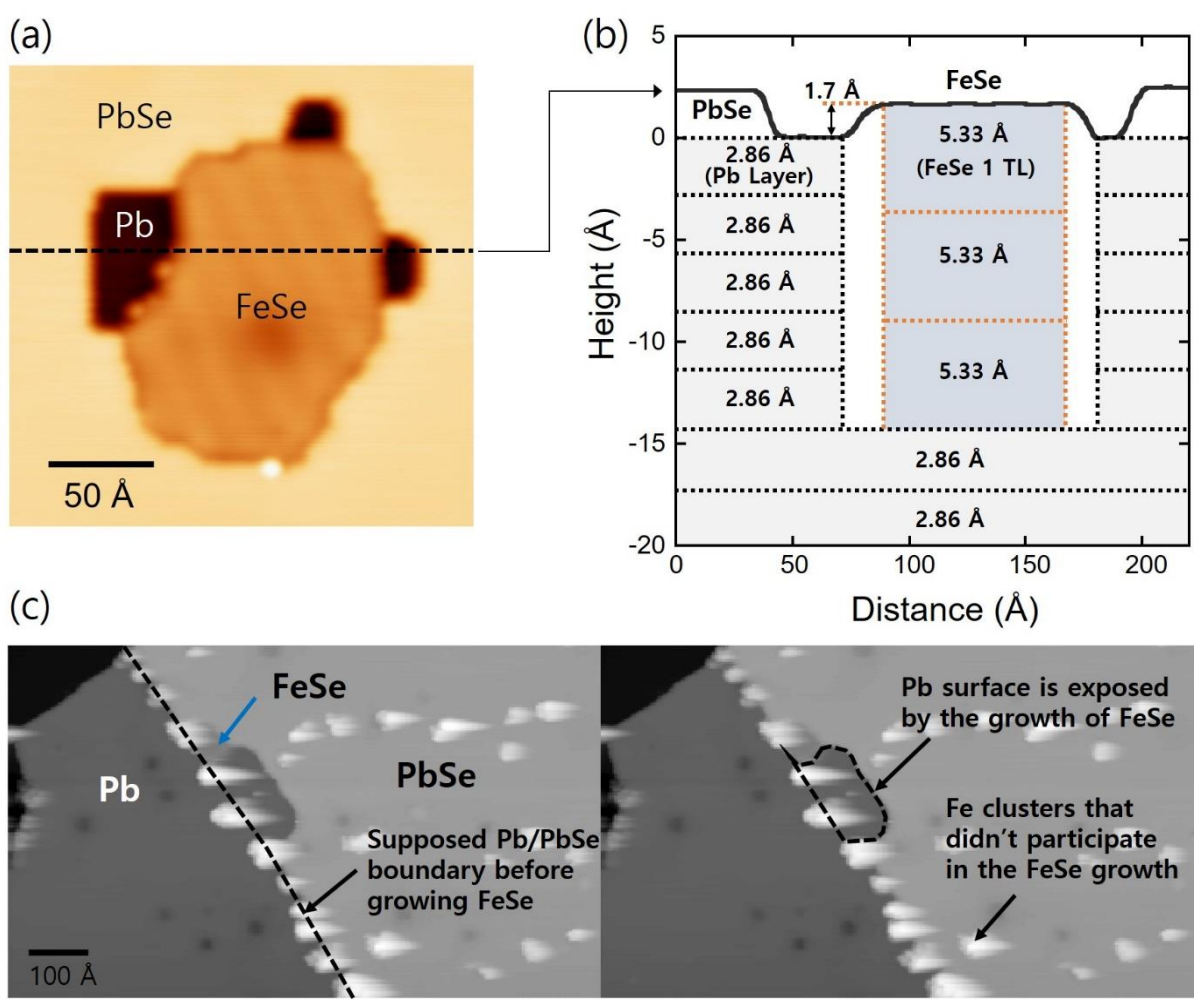

**Figure S2. Thickness of FeSe grown on Pb(111) substrate.** (a) The topography of FeSe grown on Pb(111) substrate. (b) The height profile along the dashed line in (a). The height of FeSe with respect to the Pb is found to be ~ 1.7 Å. This value corresponds to the height difference between FeSe film of 3 tri-layer (TL) and 5 layered Pb, as illustrated by sketched stacking units of FeSe and Pb (Phys. Rev. B 84, 125437 (2016); Surf. Sci. 646, 72 (2016); J. Phys.: Condens. Matter 29, 025004 (2017)). The inter-layer distance of Pb(111) is 2.86 Å. The height of 1 TL of FeSe is 5.33 Å. (c) The FeSe island grown near the edge of PbSe. The growth of FeSe induces exposure of Pb surface in the PbSe. The newly exposed Pb area is about 4 times larger than the area of FeSe island. Taking into account the Se densities of 1 mono-layer (ML) PbSe (0.08 atom/Å$^2$) and 1 tri-layer (TL) FeSe (0.14 atom/Å$^2$), we calculate the thickness of the grown FeSe film is approximately 3 TL.

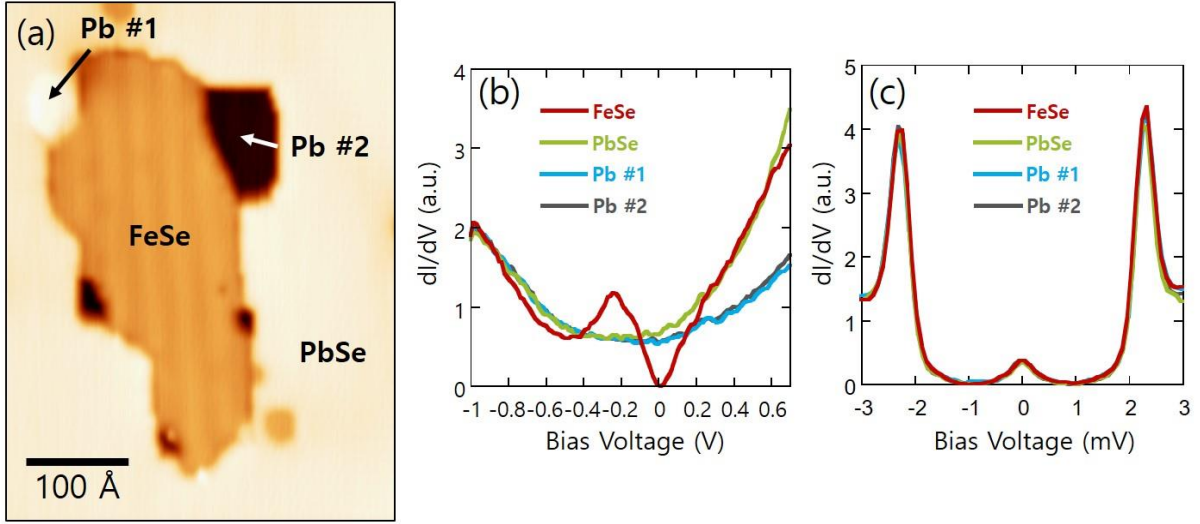

**Figure S3. Comparison among the dI/dV spectra of Pb, PbSe and FeSe.** (a) Topographic image of an FeSe grown on Pb(111). $V_{bias}$ = -0.1 V and I = 50 pA. The dI/dV spectra were taken in FeSe, PbSe and Pb surfaces. (b) The dI/dV spectra in the wide bias voltage range. $V_{bias}$ = -1 V and I = 50 pA. (c) The dI/dV spectra in the superconducting gap regime. $V_{bias}$ = - 3 mV and I = 50 pA.

| Properties | Pb | PbSe | FeSe |
|---|---|---|---|
| Origin of Superconductivity | intrinsic | proximity-Induced | proximity-Induced |
| Gap symmetry | s-wave | s-wave | s-wave |
| Superconducting Critical Temperature (Tc) | ~7.2 K | ~7.2 K | ~7.2 K (s-wave) |

**Table S1. Summary of the basic superconducting properties of Pb, PbSe and FeSe films examined in our experiment.**

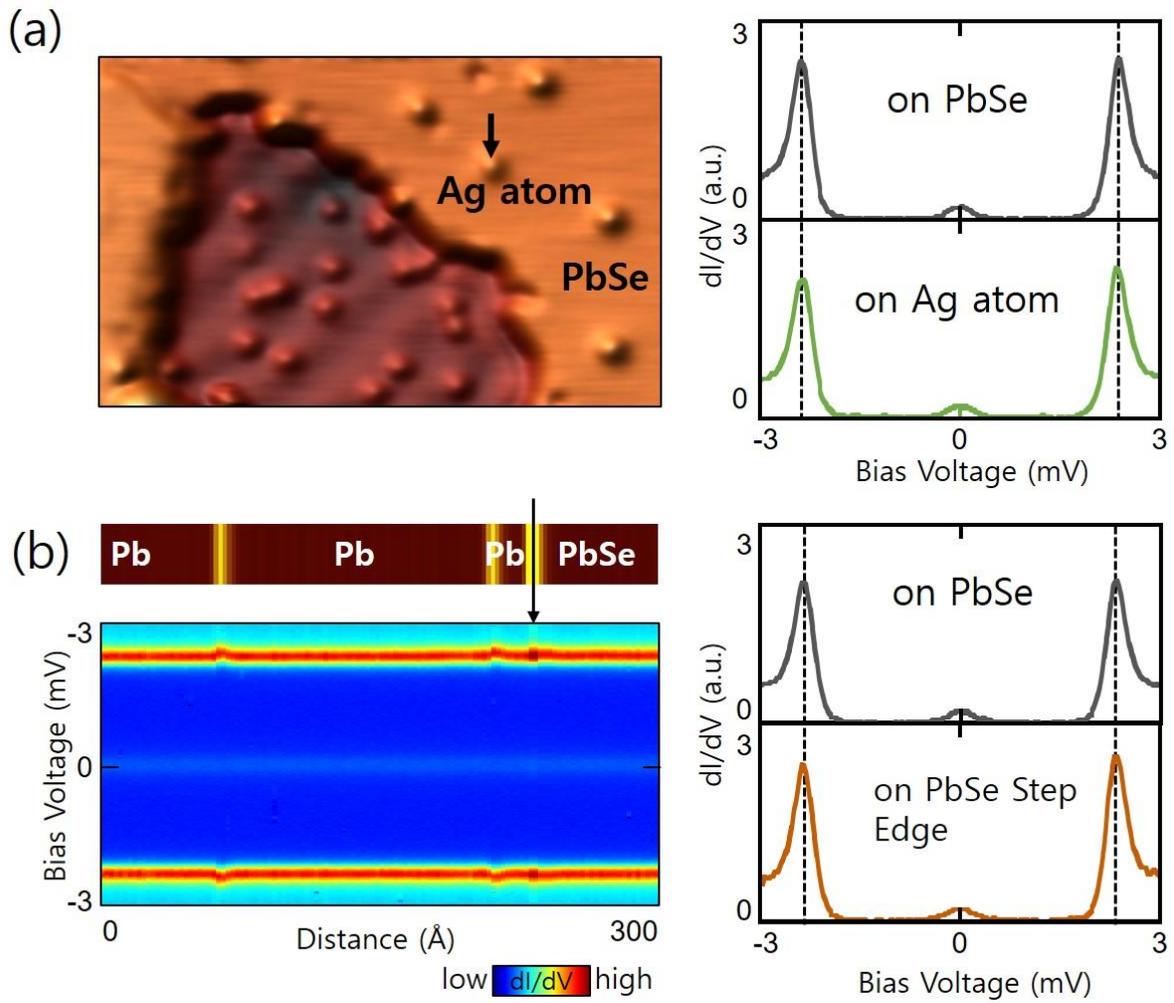

**Figure S4. Response of impurities to the proximity-induced superconductivity in PbSe.** In the main text, we show the crystal imperfections induce magnetic moments in FeSe. In this figure, we show Ag atoms and crystal edge do not induce magnetic moments in PbSe. (a) dI/dV spectra measured on/off Ag atoms on PbSe. (b) Line spectroscopy is taken across the PbSe step. The step edge is marked by the vertical arrow. No in-gap excitation is observed for the Ag atoms and the PbSe step. $V_{bias}$ = - 3 mV and I = 50 pA.

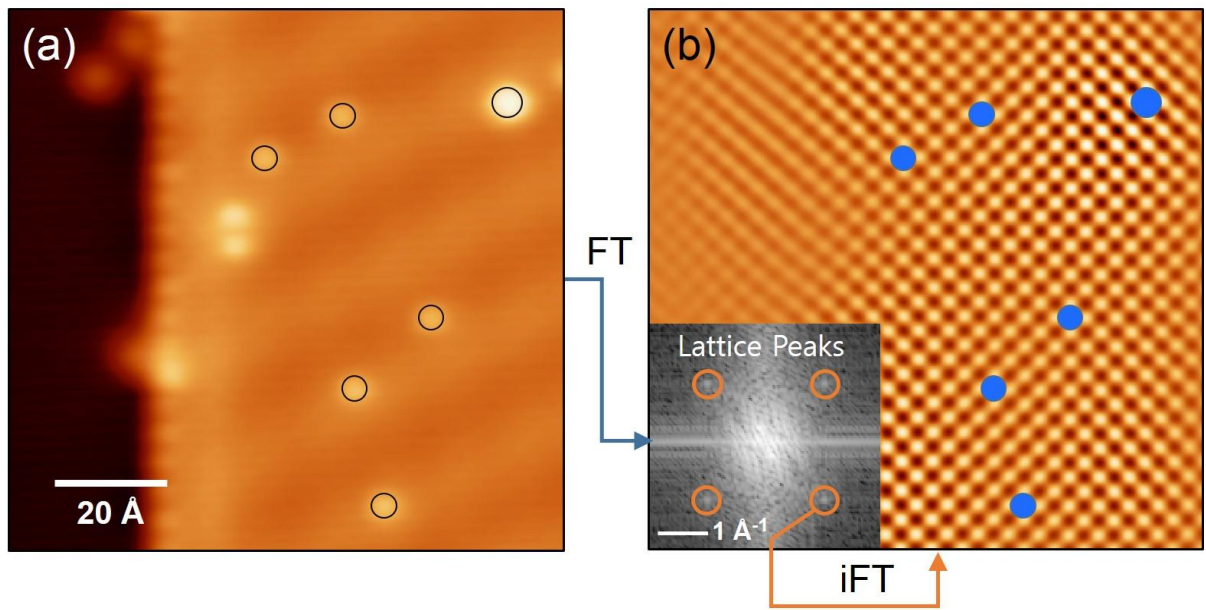

**Figure S5. Identification of the location of Ag atoms on the FeSe.** (a) Topography image of Ag atoms on the FeSe. The open circles represent the location of Ag atoms. (b) The Fourier transform image of the topography is shown in the inset. The orange circles show the lattice peaks of the top Se atoms in the FeSe. By the inverse Fourier transform of the peaks, the position of Se atoms is revealed. The blue filled circles denote the Ag atoms of which location are exactly copied from the location of the open circles in (a). The FT analysis shows the Ag atoms are located on the center of the Se lattice.

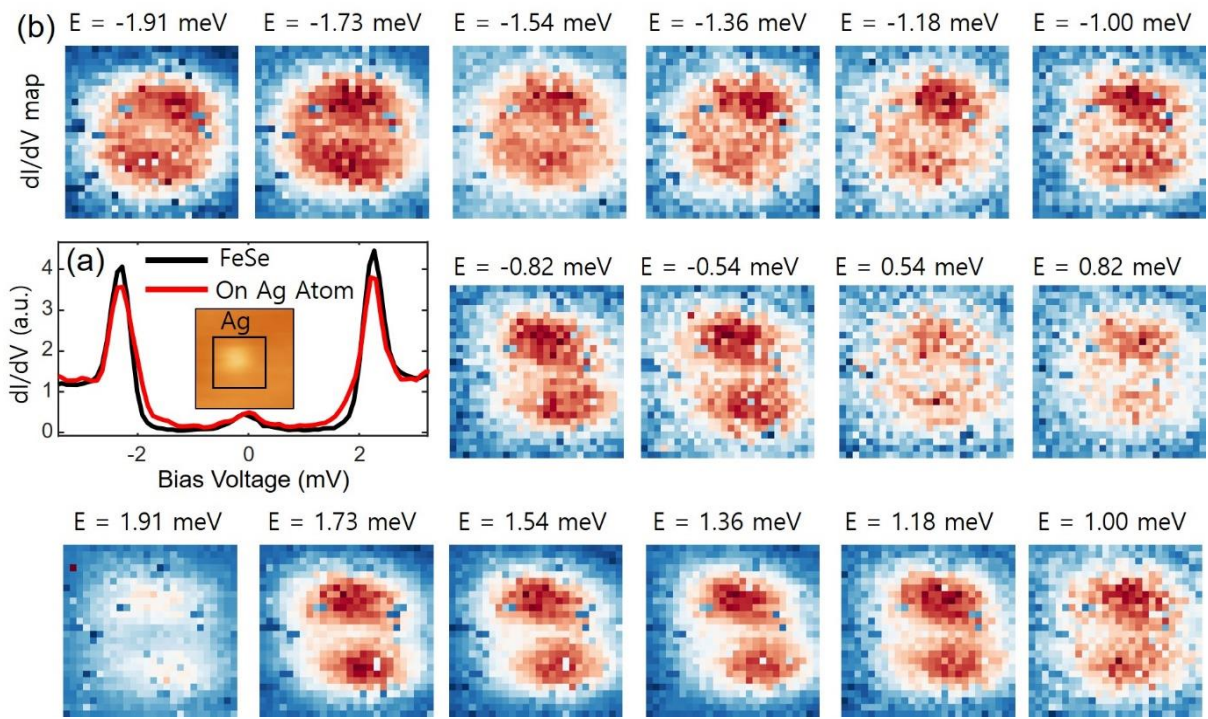

**Figure S6. Magnetic patterns induced by Ag atoms on the FeSe.** (a) The dI/dV spectra measured on FeSe (black curve) and the Ag atom (red curve). This Ag atom is different from the Ag atom in Fig. 4d and 4e. Inset shows the topography of the Ag atom. (c) The dI/dV maps for the Ag atom are displayed (the size is 8.5 Å x 8.5 Å). Every magnetic pattern shows the $C_2$ symmetry.

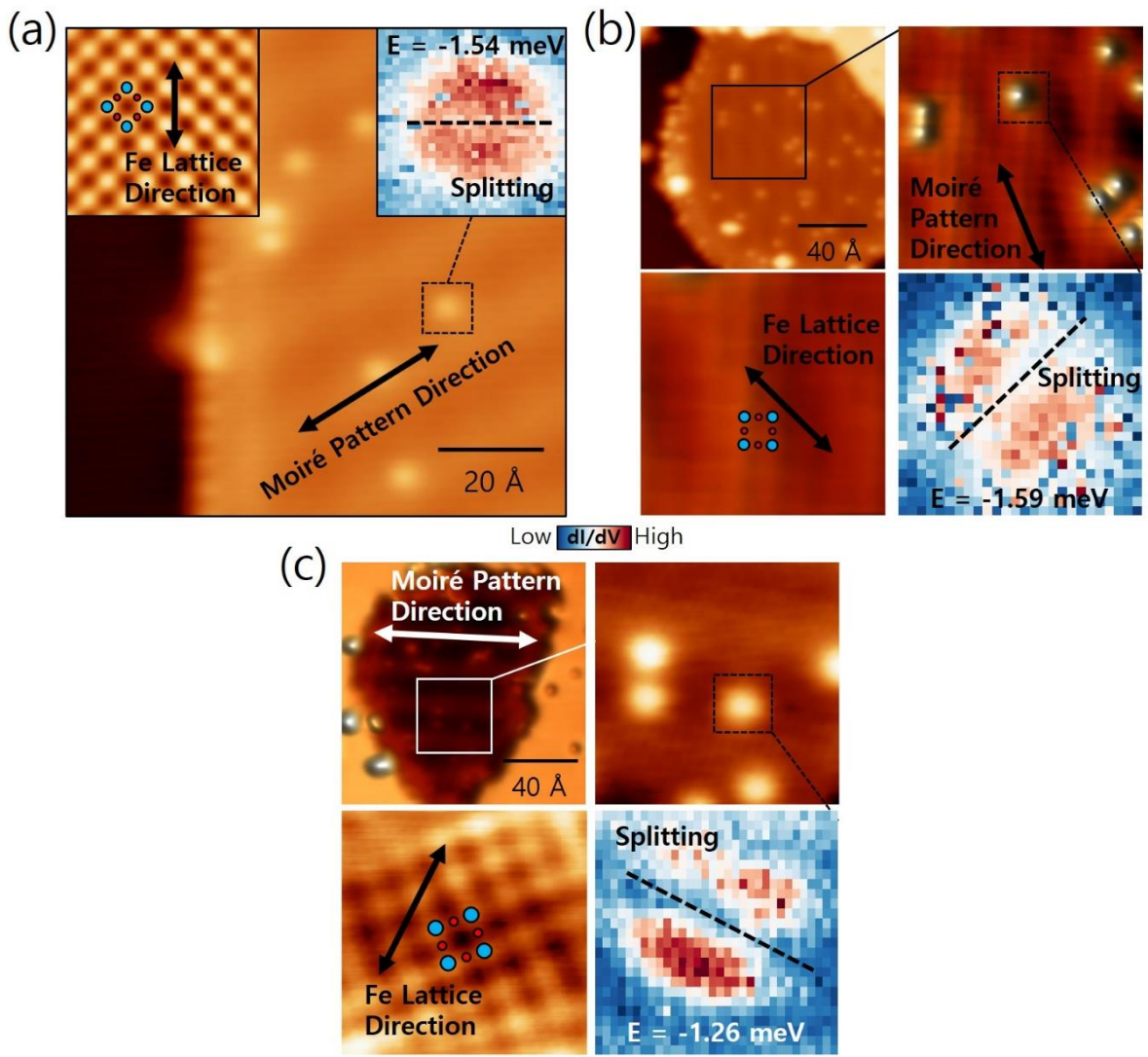

**Figure S7. Moiré pattern and the splitting pattern induced by Ag atoms in dI/dV maps.** The angle between Fe lattice and Moiré pattern is 60°, 20° and 70° for (a), (b), and (c), respectively. (a) The dI/dV map (top-right inset) is taken for the Ag atom at the energy of E = -1.54 meV. The dashed line shows the splitting of dI/dV intensity. The atomic structure (top-left inset) is obtained by Fourier transform analysis. The blue balls and red balls represent the Se lattice and Fe lattice, respectively. (b) The dI/dV map is taken at the energy of E = -1.59 meV. (c) The dI/dV map is taken at the energy of E = -1.26 meV. Regardless of the Moiré pattern, the splitting in the dI/dV maps is aligned along the Fe lattice direction, which rules out the Moiré pattern as a possible origin of the splitting in the dI/dV maps. The data shown here were obtained using different tips in the separate experiments.

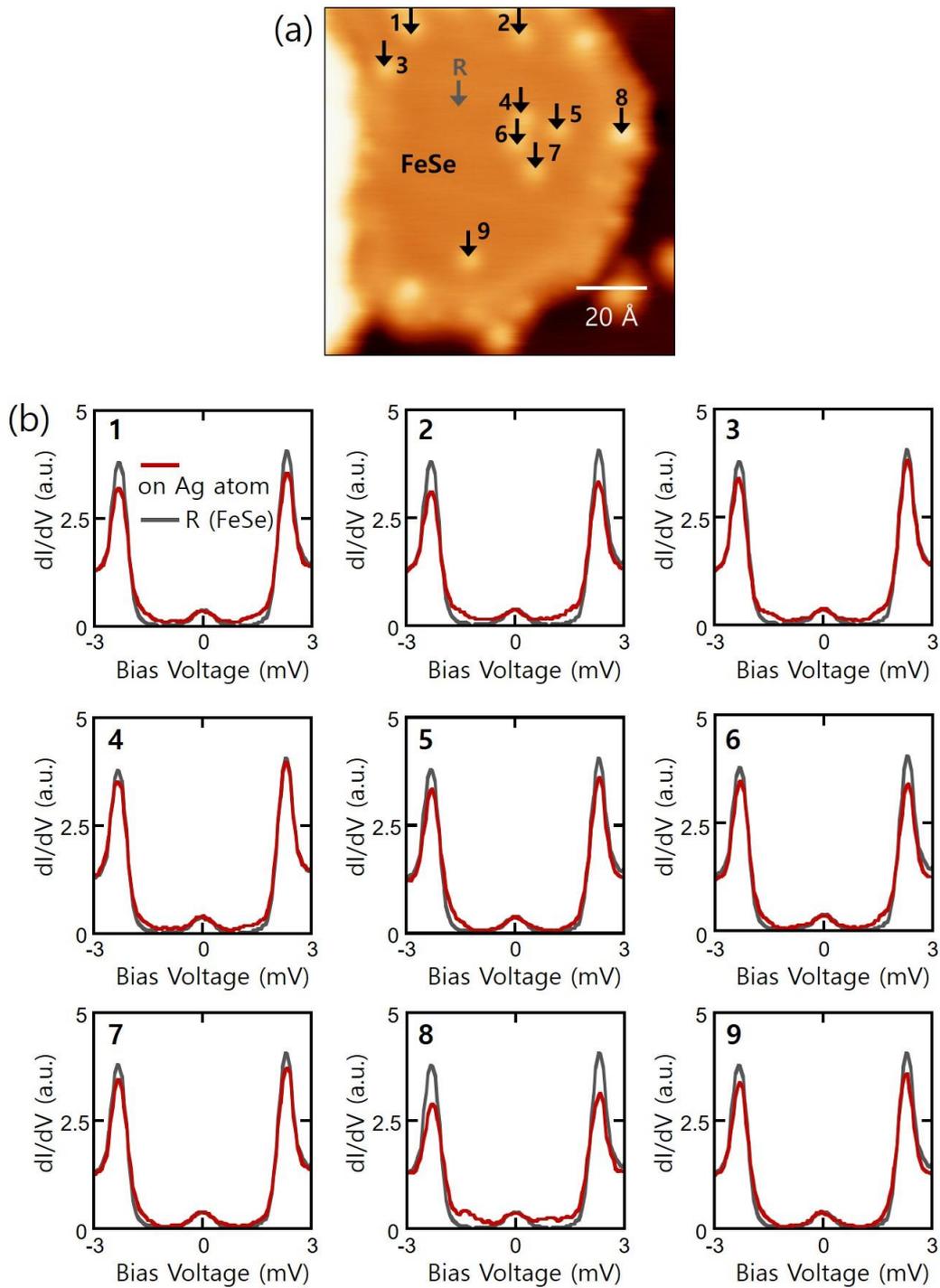

**Figure S8. dI/dV spectra measured on Ag atoms in FeSe.** (a) Ag atoms on the FeSe island. Imaging condition: $V_{bias}$ = -0.1 V and I = 50 pA. (b) The dI/dV spectrum for each numbered Ag atom in (a) is displayed. The spectrum measured on the Ag atom show YSR excitation compared to the spectrum measured on FeSe. The spectroscopy condition: $V_{bias}$ = 3 mV and I = 50 pA. Lock-in modulation: f = 463.0 Hz and $V_{rms}$ = 60 μV.

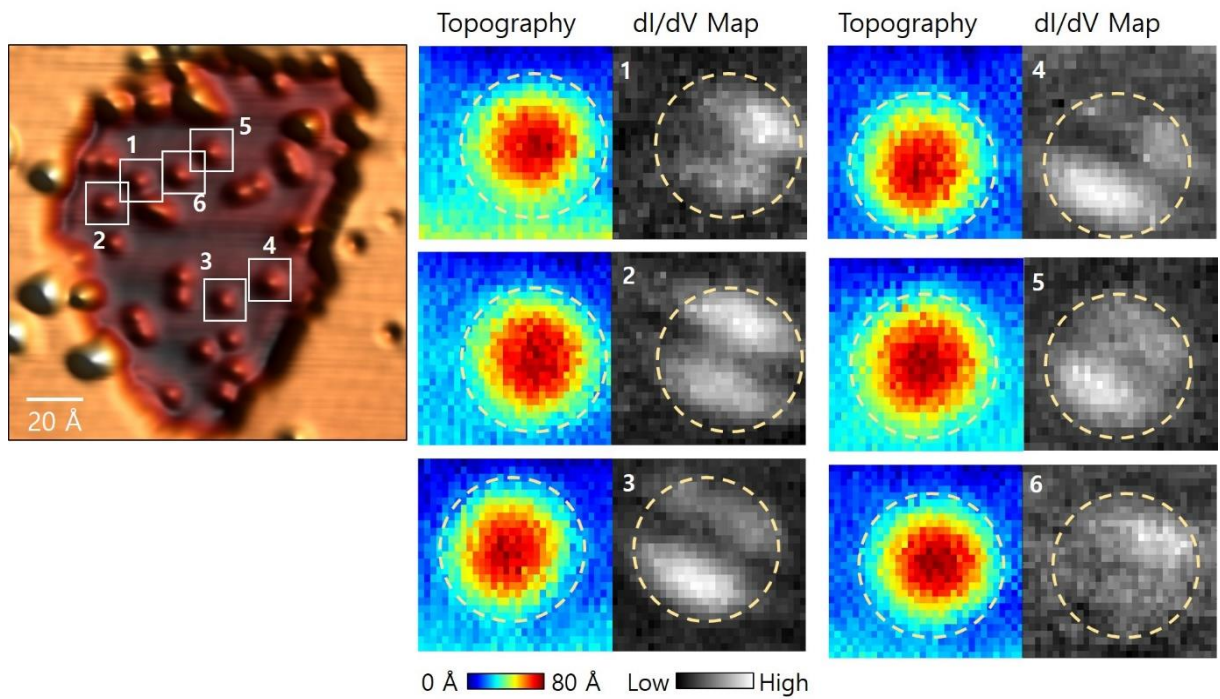

**Figure S9. Magnetic patterns for various Ag atoms on the FeSe.** We measured dI/dV maps at E = -1.1 meV for isolated Ag atoms. The studied Ag atoms are indicated with numbers in the left panel. The topography and simultaneously obtained dI/dV map are displayed in the right panels. Most of Ag atoms exhibit the splitting patterns in dI/dV maps except for the Ag atom labeled with 6. The variation could be because the Ag atoms are not perfectly positioned at the center of Fe lattice.

| Spin Angle | Collinear AFM | Néel AFM |
|---|---|---|
| 45° | | |
| 0° | | |

**Figure S10. Symmetry of Collinear AFM and Néel AFM models with spin angle of 0°.** The collinear AFM model with the spin angle of 0° or 45° is fully consistent with the measured dI/dV maps in Fig. 3e in the main text in terms of symmetry. The Néel AFM model with the spin angle of 0° does not preserve $C_2$ symmetry, which is inconsistent with the dI/dV maps. Furthermore, it maintains a mirror symmetry as indicated by the dashed line in the image, which contradicts the symmetry of the dI/dV map at E = 1.48 meV. Therefore, the Néel AFM model does not explain the symmetry of the local magnetic moments induced by the Ag impurities in the experiment.

**Note S2. Theory results for local magnetic order around Se centered impurities and FeSe island edges.**

In this section we expand the theoretical study of local magnetic order around Fe-centered impurity bound states in Ref. S11A to also include Se-centered disorder and edges. These calculations are performed in a similar fashion to that described in detail in Ref. S11A, and therefore we provide only a brief outline here.

We perform self-consistent mean-field calculations in the Hubbard-Hund model using the tight-binding parameters for FeSe described in Ref. S11B. The interaction parameters are fixed in terms of the Hubbard U as $J = J' = U/4$ and $U' = U - 2J$.

Orbital selective effects are included by a rescaling of electron creation and annihilation operators $c_\mu \to \sqrt{Z_\mu} c_\mu$, with $Z_\mu$ the quasiparticle weight factor in the given orbital, yielding an effective model with rescaled orbital-dependent interaction parameters

$$U_{\mu\nu} \to \sqrt{Z_\mu}\sqrt{Z_\nu} U_{\mu\nu},$$

with similar expressions for $U'$, $J$, $J'$. Based on Ref. S11B. we choose these weights as $\sqrt{Z_\mu} = 0.2715, 0.9717, 0.4048, 0.9236, 0.5916$ for the five Fe 3d orbitals $\mu = d_{xy}, d_{x^2-y^2}, d_{xz}, d_{yz}, d_{z^2}$.

In this model a phase transition to a strongly $C_2$-symmetric magnetically ordered bulk phase occurs at a critical $U_c = 560$ meV. As the Hubbard U approaches this transition from below, it was previously found that impurity bound states may facilitate local magnetic order [S11A]. In the following we thus fix $U = 550$ meV just below the critical value, but remark that our results in general apply to an interval of $U$, the width of which depends on the type of disorder or impurity potential.

Finally, large-scale real-space calculations are facilitated by employing the Kernel Polynomial method where the electronic Greens function is expanded in a series of orthonormal Chebyshev polynomials [S11C, S11D]. We set the order of this expansion to $N = 1000$, use the Lorentz kernel to damp Gibbs oscillations, and iterate self-consistently until convergence of the spin resolved density mean fields $(n_\uparrow, n_\downarrow)$ is obtained.

We model the Se vacancy as an effective plaquette impurity, i.e. by an onsite potential V on the four neighboring Fe sites of the vacancy on a given site

$$H_{Se-imp} = V_{Se} \sum_{j,\mu,\sigma} c^\dagger_{j\mu\sigma} c_{j\mu\sigma},$$

where $V_{Se}$ is the potential applied to the four neighboring Fe sites indexed by $j$. Calculations of the induced local magnetic order are performed using such a plaquette impurity in the center of a 12 x 12 supercell.

Our studies of edge magnetism are performed using open boundary conditions in the real-space system. This creates an isolated FeSe island with boundaries determined by the system geometry. In our calculations we modify the original periodic supercell structure to accommodate both (100), (010) and (110) edges on the islands.

Fig. S11 (a-b) displays the result of including a single central plaquette impurity for $U$ close to the phase boundary. In Fig. S11 (a) we show a zoomed-in real-space plot of the magnetization centered on the impurity site, while Fig. S11 (b) displays the associated 2D Fourier transform. We find that the plaquette impurity induces local magnetic order for a broad range of impurity potentials $V_{Se}$. Similar to the point-like impurity, we find that the local magnetic order inherits the structure of the bulk magnetic fluctuations, yielding a strongly non-$C_4$-symmetric structure of the induced magnetization, as demonstrated by the peaks in the Fourier transform at $m_z(q) = (\pm \pi, 0)$.

In Fig. S11 (c-f) we show results for the local magnetic order formed on FeSe island edges when open boundary conditions are imposed in the calculation. In general, we find that as U is increased from below, magnetic order forms initially on corners of the system, but that closer to the phase boundary magnetization is induced along the entire edge of the system. This is demonstrated for a geometry with long 100 edges in (c) and for 110-type edges in the different geometry in (d). We note that the included orbital selective effects also make the amplitude of the induced magnetization at 100 (extending along x) versus 010 (extending along y) edges distinct.

Fig S11 (e) shows the linecut of the magnetization in (c) indicated by the dashed grey line transverse to the 100 edge. The magnetization peaks sharply directly on the edge, but a tail of finite magnetization extends into the bulk. In (f) we show two linecuts taken transverse to the edge through the magnetization of the FeSe island with 110-type edges plotted in Fig. S11 (d). For the staggered 110 edge, the magnetization selectively forms on every second site of the edge, forming a chain conforming to the bulk magnetic order. The linecuts taken transverse to the edge at two neighboring edge sites demonstrates this feature. The linecuts are mirror images of each other, the dashed cut showing the peak in magnetization at the upper edge with an oscillating tail into the bulk, and the cut corresponding to the dotted line has the opposite structure with a peak at the lower edge.

[S11A]    J.H.J. Martiny, A. Kreisel, and B.M. Andersen, Physical Review B **99**, 014509 (2019)
[S11B]    A. Kreisel *et al.*, Physical Review B **95**, 174504 (2017)
[S11C]    A. Weiße, G. Wellein, A. Alvermann, and H. Fehske, Rev. Mod. Phys. **78**, 275 (2006).
[S11D]    L.Covaci, F.M. Peeters, and M. Berciu, Phys. Rev. Lett. **105**, 167006 (2010).

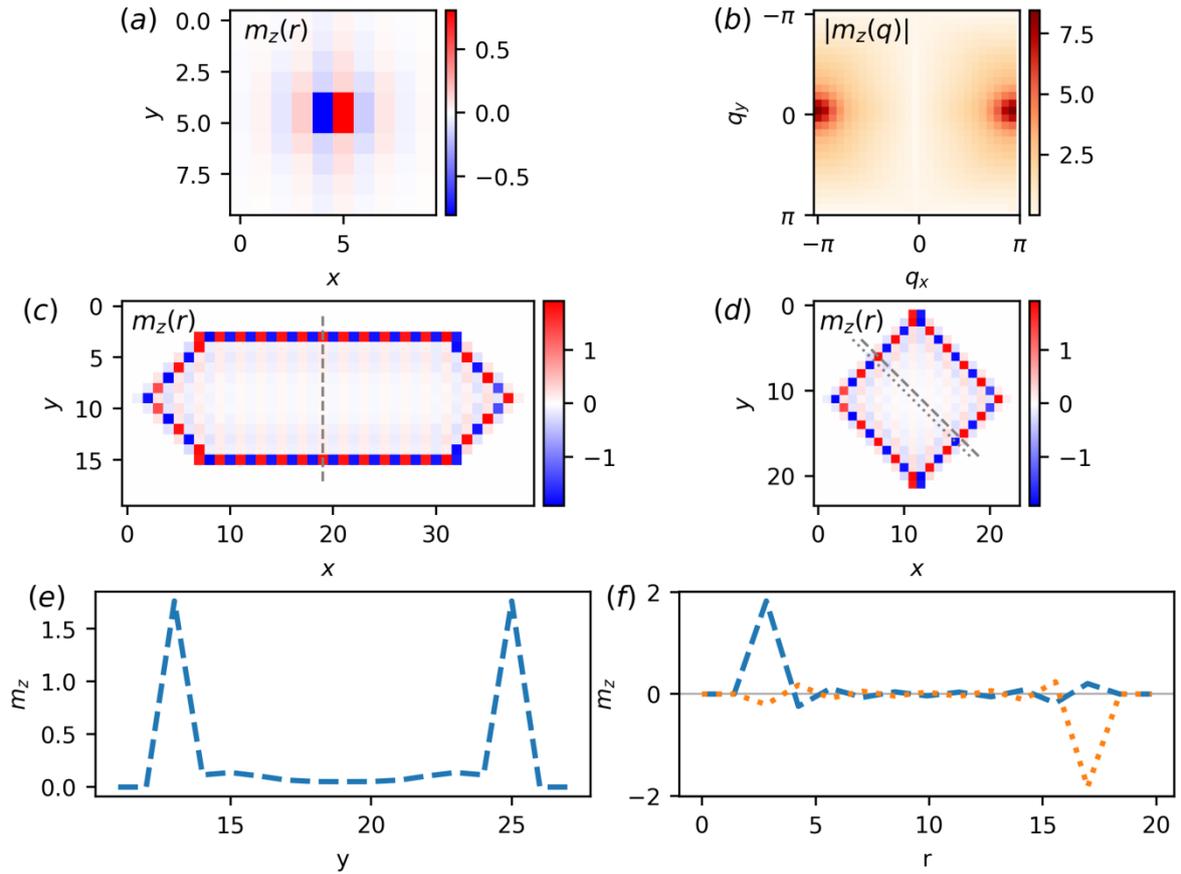

**Figure S11. Theoretical results for local magnetic order near impurity sites and edges.** (a) Zoom of the magnetization nucleated around a plaquette impurity $V_{Se} = 50$ meV. (b) Fourier transform of (a) showing the $C_2$ structure of the local magnetic order. (c) Edge magnetization on a FeSe island with open boundary conditions and a long 100 edge. (d) Edge magnetization on a different FeSe island with 110-type edges. (e) Linecut of the magnetization for the 100 edge [dashed line in (c)]. (f) Linecuts of the magnetization for the 110 edge [dashed, dotted lines in (d)]. In both geometries the magnetization peaks at the edge with a tail extending into the bulk region.

**Note S3. The verification of the experimental results using Au adatoms on the FeSe.**

To check if any other impurities induce similar local magnetic moments on the FeSe as the Ag adatoms do, we evaporated gold (Au) adatoms on the FeSe (Fig. S12). We observed the Au adatoms induce strong YSR bound states in the FeSe, which further confirms our conclusion in the manuscript that the ground state of FeSe is close to the magnetic quantum critical point.

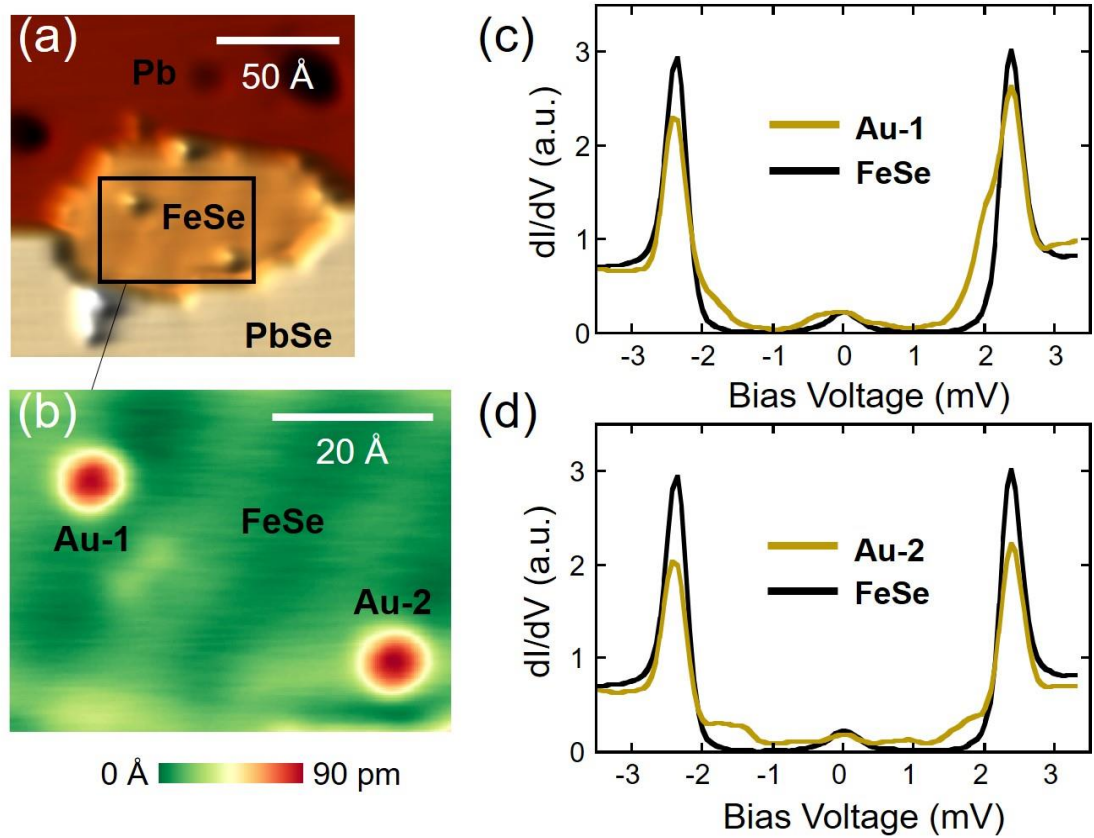

**Figure S12. The local magnetic moments induced by Au atoms on the FeSe.** (a) FeSe film grown on Pb(111) substrate. The Au adatoms are deposited on the FeSe film. (b) Zoomed-in image of the FeSe Film. (c, d) The YSR bound states are clearly seen on the Au adatoms, implying that the magnetic moments are induced by Au atoms in FeSe. The spectroscopy condition: $V_{bias}$ = 3.5 mV and I = 50 pA. Lock-in modulation: f = 463.0 Hz and $V_{rms}$ = 60 μV.